\input{aipcheck}

\documentclass[
    ,final            
  ]
  {aipproc}

\layoutstyle{8x11single}

\newcommand{\etal}{et~al.\ }

\newcommand{\be}{\begin{equation}}
\newcommand{\ee}{\end{equation}}
\newcommand{\ba}{\begin{eqnarray}}
\newcommand{\ea}{\end{eqnarray}}

\newcommand{\swift}{\emph{Swift}}

\begin{document}

\title{Late-Time Optical Afterglow Observations with LBT and MDM}

\classification{98.70.Rz}
\keywords      {Gamma-ray bursts}

\author{X. Dai}{
  address={Department of Astronomy, Ohio State University, Columbus, OH 43210, USA}
}

\author{K. Z. Stanek}{
  address={Department of Astronomy, Ohio State University, Columbus, OH 43210, USA}
}

\author{P. M. Garnavich}{
  address={Department of Physics, University of Notre Dame, Notre Dame, IN 46556, USA}
}

\begin{abstract}
Using the 2.4m MDM and 8.4m Large Binocular Telescope, we observed nine GRB afterglows to systematically probe the late time
behaviors of afterglows including jet breaks, flares, and supernova bumps.  In particular, the LBT observations have typical
flux limits of 25-26 mag in the Sloan $r'$ band, which allows us to extend the temporal baseline for measuring jet breaks by
another decade in time scale. We detected four jet breaks (including a ``textbook'' jet break in GRB070125) and a fifth
candidate, all of which are not detectable without deep, late time optical observations. In the other four cases, we do not
detect the jet breaks either because of contamination from the host galaxy light, the presence of a supernova bump, or the
intrinsic faintness of the optical afterglow. This suggests that the basic picture that GRBs are collimated is still valid and
that the apparent lack of Swift jet breaks is due to poorly sampled afterglow light curves, particularly at late times.
Besides the jet breaks, we also detected late time flares, which could attribute to
late central engine activities, and two supernova bumps.
\end{abstract}

\maketitle


\section{Introduction}
Despite the tremendous success of the \swift\ mission (Gehrels et al.\ 2004), 
the late time optical afterglow observations still remain an area seldom 
explored by observers.
The deep, late time observations are important to constraint the jet break time, 
detect supernova bumps, flares, and emission from the GRB host galaxies. 
All of these features have been detected before \swift, such as the jet break
in GRB~990510 (Stanek et al.\ 1999) and the supernova bump in GRB011121 
(Garnavich et al.\ 2003).  

After the launch of \swift, with its rapid localization of GRBs and the dedicated on-board XRT instrument, the number of GRBs with jet breaks was expected to increase significantly.
Instead, there are few jet break detected, and there arise a ``missing jet break'' problem.
However, the complicated decay patterns both in X-rays and the optical make it
more difficult to identify a jet break.
While the flares are complications, the biggest problem for identifying
jet breaks is the need for well-sampled, long term light curves.
While almost all \swift\ bursts have good signal-to-noise ratio (S/N) X-ray light curves at early decay times,
most of the jet breaks occur at late times when the uncertainties in the XRT light curves increase significantly.  In these cases, the claim that the data are consistent with the single power-law decay does not distinguish between models, because the error-bars are so large that the data are also consistent with the broken power-law of a jet break.
In the optical bands, the afterglow monitoring is generally sparse, in part 
because of the large number of bursts detected.  
Thus, the number of \swift\ bursts with jet breaks or supernova bumps is small.

Given the need for deep, late-time monitoring of GRBs, we initiated an observing program using the MDM 2.4 meter telescope and the newly built 8.4 meter Large Binocular Telescope (LBT, Hill et al. 2006).  
In particular, with 20--30 minute exposures, the LBT can reach a flux limit of $\sim25$--$26$~mag in the Sloan $r'$ band.  
This allows us to extend the temporal baselines of optical afterglow monitoring by roughly an order of magnitude, and systematically search for the jet breaks and other emission components in the optical light curves.

\section{MDM and LBT Observations}
We observed 3 GRBs, GRB~060206, GRB~060210, and GRB~060526 with the MDM 2.4$m$ 
telescope in the optical $R$ band in February and May 2006.
We also observed 6 GRBs, GRB~070125, GRB~070311, GRB~070411, GRB070412, 
GRB070419A, GRB070518, with the 8.4$m$ LBT using the Large Binocular Camera 
(LBC) Blue CCD camera (Ragazzoni et al. 2006) during the LBT/LBC Science 
Demonstration Time period in January--November 2007.

\section{Results}

\begin{table}
\begin{tabular}{lccccc}
\hline
  & \tablehead{1}{c}{b}{Telescope}
  & \tablehead{1}{c}{b}{$R$ Limit (mag)}
  & \tablehead{1}{c}{b}{Optical Break?}
  & \tablehead{1}{c}{b}{X-ray Break?}
  & \tablehead{1}{c}{b}{Simple Jet Model?}   \\
\hline
060206 & MDM & 21.5 & yes & consistent & yes \\
060210 & MDM & 21.5 & faint & &  \\
060526 & MDM & 23.5 & yes & consistent & no \\
070125 & LBT & 25--26 & yes & yes & yes \\
070311 & LBT & 25--26 & yes & consistent & no \\
070411 & LBT & 25--26 & candidate & consistent &  \\
070412 & LBT & 25--26 & faint & &  \\
070419A & LBT & 25--26 & contamination & & \\
070518 & LBT & 25--26 & contamination & &  \\
\hline
\end{tabular}
\caption{Jet Search Summary}
\label{tab:one}
\end{table}

\begin{figure}
  \includegraphics[height=.3\textheight]{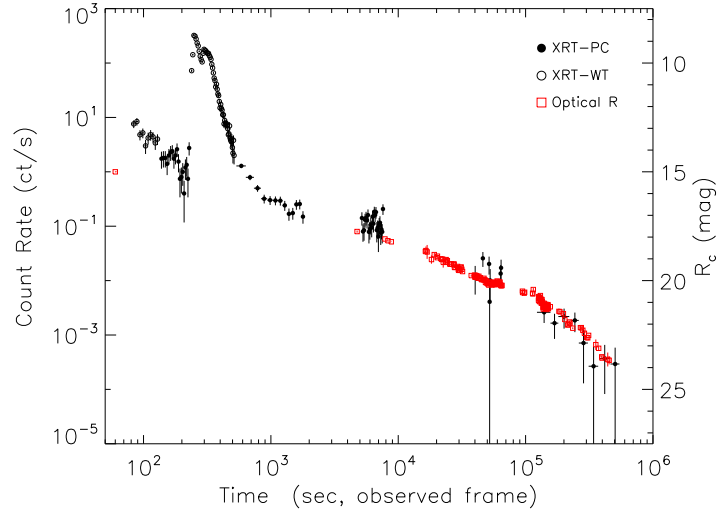}
  \caption{X-ray and optical afterglow light curves of GRB~060526.  The open and filled circles are data from the XRT Windowed Timing mode and Photon Counting mode, respectively, and the squares are the R band optical data from our MDM 1.3m, 2.4m, and PROMPT observations combined with other optical observations reported in the GCN. 
We observed a break at $2.4\times10^5$~sec in the optical afterglow light curve (Dai et al.\ 2007).
\label{fig:lc}}
\end{figure}

\begin{figure}
  \includegraphics[height=.3\textheight]{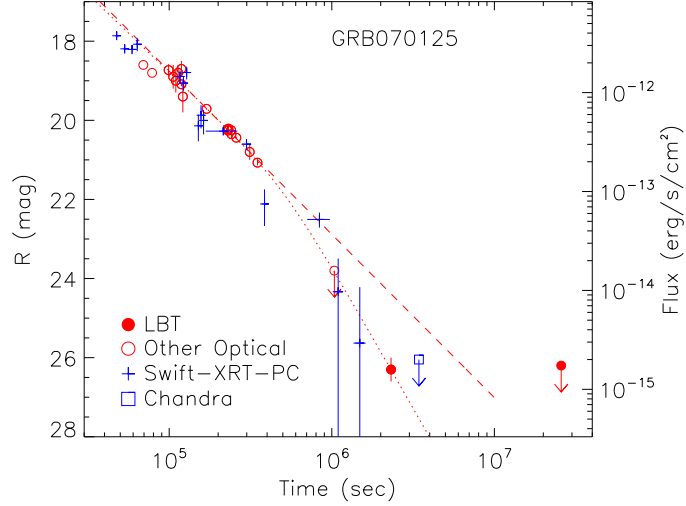}
  \caption{Optical and X-ray light curves of GRB~070125 (Dai et al.\ 2008).  
The dashed line is a single power-law fit ($\alpha=1.65\pm0.05$) to the optical data from $9\times10^4$ to $4\times10^5$~s.
The dotted line is a broken power-law fit to the whole optical data set with $\alpha_1=1.58\pm0.12$, $\alpha_2=2.87\pm0.17$, and $T_b = (5.0\pm1.6)\times10^5$~s.
The X-ray and optical data are aligned such that the steepest single power fit to the X-ray data ($\alpha_X = 1.7$) overlaps the single power fit to the optical data (the dashed line).
\label{fig:lcone}}
\end{figure}

\subsection{Jet Breaks}
We summarize our jet search results in Table~1.
In the sample of nine bursts we observed, we detected four optical jet breaks 
and a fifth candidate (Stanek et al.\ 2007; Dai et al.\ 2007, 2008).
We show two examples GRB~060526 and GRB~070125 in Figures~1 and 2.
In the other four cases, the jet break cannot be detected either due to the
intrinsic faintness of the afterglow or the presence of additional contamination 
from host galaxies of supernova bumps.
In the five bursts, where we detected an optical jet break/candidate, the X-ray
light curves of these bursts are all consistent with the optical light curves,
except for GRB070311, where optical and X-ray light curves are only consistent 
at post-break decay stage.  In GRB~070125, a break is independently required from
the X-ray light curve alone.  This indicates all the break we detected 
are achromatic, and consistent with predictions from the jet models.
However, if we only examine the X-ray data, only one break can be detected.
Likewise, if we only monitor the optical afterglow to about 21--22 magnitude,
we would only find one break.
The results suggest that late time optical observations are the key to detect
jet breaks.

When we further confront the jet model with the predictions from spectral and
temporal decay slopes before and after the breaks, we found two bursts
satisfy the prediction from simple jet models.  In particular, GRB~070125 shows a ``textbook jet break'' (Dai et al.\ 2008).
The break is independently seen in both optical and X-ray light curves,
and the consistency of the optical and X-ray light curves suggests that the break is achromatic.
Furthermore, the spectral and temporal decay indices before and after the break satisfy the predictions from the standard jet models.
This result is confirm by an independent analysis (Updike et al.\ 2008).
There are still two bursts, where temporal and spectral slopes do not satisfy predictions from the simple jet model.  This could be caused by additional components in the
afterglow light curves, or that the jet model is more complex (e.g., Panaitescu et al.\ 2006).

Overall, the results suggest that the basic picture that GRBs are collimated is still valid and that the apparent lack of Swift jet breaks is due to poorly sampled afterglow light curves, particularly at late times.

\subsection{Flares}
We observed significant optical flares both in early (GRB~060206) and late time 
(GRB~060526 and GRB~070311) decay stages, indicating the optical afterglows 
decay patterns are also complex.  For the late time flares in GRB~060526 and 
GRB~070311, we found the relative amplitudes of the flares are large 
($\Delta F/F = 0.2, 0.7, 0.2, 2.5$), and they only last in a relatively short 
time ($\Delta T/T = 0.23, 0.75, 0.22, 2$).  These observations, with $(\Delta F/F)/(\Delta T/T) \sim 1$, are not consistent
with variability caused by ambient density fluctuations (Ioka et al.\ 2005),
 and it is possible that these flares are associated with late central energy activities.  We show the late flares in GRB~060526 and GRB~070311 in Figures~3 and 4.

\begin{figure}
  \includegraphics[height=.3\textheight]{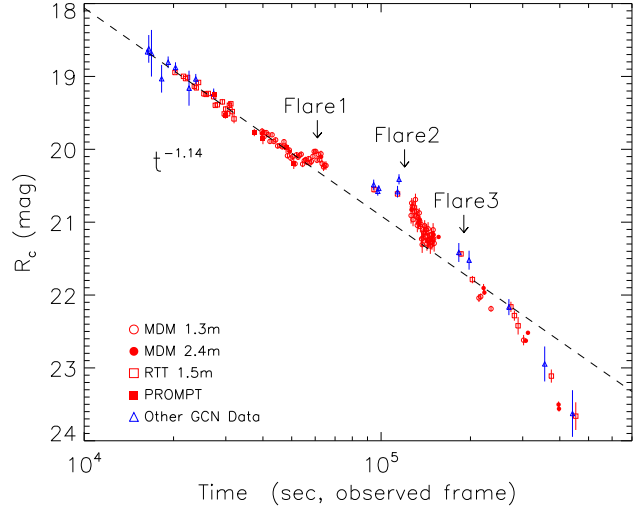}
  \caption{The densely sampled optical light curve from $10^4$~sec after the burst trigger for GRB~060526.  The open circles, filled circles, and fill squares are our MDM 1.3m, 2.4m, and PROMPT observations and the rest of the data are from GCN circulars.  
The data before $5.4\times10^4$~sec is fitted by a power-law with a slope of $1.14\pm0.02$ (dashed line).  Compared with the single power-law fit, the late optical data clearly show multiple flares and a much steeper late time decay slope of $3.4\pm0.2$ (Dai et al.\ 2007).\label{fig:opt}}
\end{figure}

\begin{figure}
  \includegraphics[height=.3\textheight]{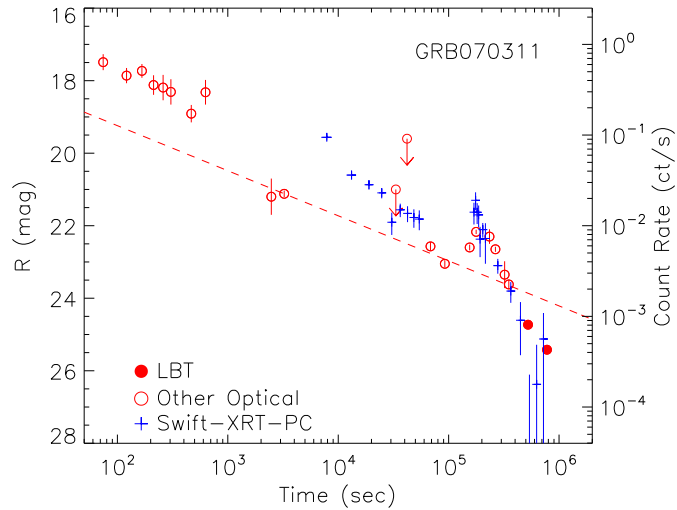}
  \caption{Optical and X-ray light curves of GRB~070311.  The late-time optical light curve contains a significant flare at $2\times10^5$~s. (Dai et al.\ 2008)}
\end{figure}

\subsection{Supernova Bumps}
We also detected supernova bumps in GRB~070419A and GRB~070518.  In Figure~5, we show
the supernova bump in GRB~070518 as an example.  The detailed analysis will be presented in Garnavich et al.\ (2008, in preparation) and McClelland et al.\ (2008 in preparation).

\begin{figure}
  \includegraphics[height=.4\textheight]{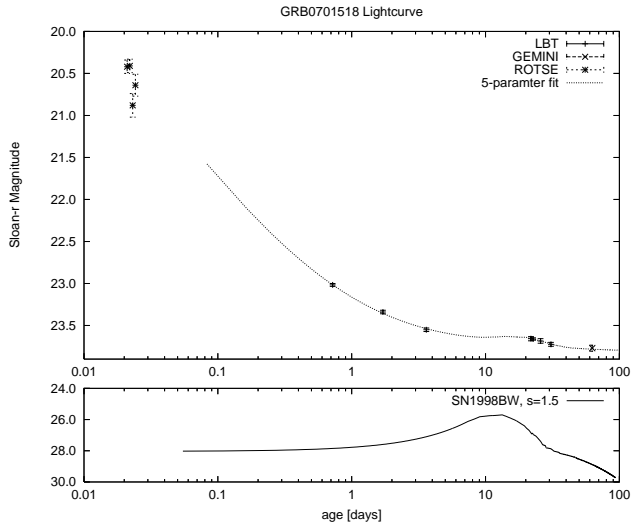}
  \caption{Optical light curve of GRB~070518 fit with three components, the afterglow, supernova bump, and host galaxy (McClelland et al.\ 2008 in preparation).}
\end{figure}

\section{Summary}
Deep, late-time optical monitoring is essential to search for jet breaks.
Our current jet search results suggest that the basic picture that GRBs are collimated is still valid and
that the apparent lack of Swift jet breaks is due to poorly sampled afterglow light curves, particularly at late times.
Searching only in the X-ray band or from shallow optical light curves will miss a significant portion of jet breaks.
Invalidating the beaming model needs well-sampled light curves.
Besides the jet breaks, we also detected late time flares, which could attribute to
late central engine activities, and two supernova bumps.


\begin{theacknowledgments}
We thank our collaborators J.~L.~Prieto, N.~D.~Morgan, C.~McClellend, J.~P.~Halpern, E.~Armstrong, N.~Mirabal, J.~B.~Haislip, D.~E.~Reichart, and the LBT collaboration.
We acknowledge the \swift\ team for the prompt detection and localization of the GRB and the rapid release of data products.  We also thank the GRB Coordinates Network (GCN) and astronomers who contribute to the GCN circular.
\end{theacknowledgments}



\bibliographystyle{aipprocl} 


\begin{thebibliography}{}

\bibitem[Dai et al.(2007)]{2007ApJ...658..509D} Dai, X., et al.\ 2007, ApJ, 658, 509 

\bibitem[Dai et al.(2008)]{2008ApJ...682L..77D} Dai, X., et al.\ 2008, ApJ, 682, L77 

\bibitem[Garnavich et al.(2003)]{2003ApJ...582..924G} Garnavich, P.~M., et 
al.\ 2003, ApJ, 582, 924

\bibitem[Gehrels et al.(2004)]{gehrels04} Gehrels, N., et al.\ 
2004, ApJ, 611, 1005 

\bibitem[Hill et al.(2006)]{hill06} Hill, J.~M., Green, R.~F., \&
Slagle, J.~H.\ 2006, Proc. SPIE, 6267, 62670Y

\bibitem[Ioka et al.(2005)]{ioka05} Ioka, K., Kobayashi, S., \& Zhang, B.\ 2005, ApJ, 631, 429 

\bibitem[Panaitescu \etal (2006)]{panaitescu06}Panaitescu, A., 
M{\'e}sz{\'a}ros, P., et al.\ 2006, MNRAS, 369, 2059

\bibitem[Ragazzoni et al.(2006)]{rag06} Ragazzoni, R., et al.\ 2006,
Proc. SPIE, 6267, 626710

\bibitem[Stanek \etal(2007)]{stanek07} Stanek, K. Z., et al.\ 2007, ApJ, 654, L21

\bibitem[Stanek et al.(1999)]{stanek99} Stanek, K.~Z., Garnavich, P.~M., et al.\ 1999, ApJ, 522, L39 

\bibitem[Updike et al.(2008)]{updike08} Updike, A.~C., et al.\ 2008, ApJ, 685, 361

\end{thebibliography}



\end{document}